# Simple Muscle Architecture Analysis (SMA): an ImageJ macro tool to automate measurements in B-mode ultrasound scans.

Olivier R. Seynnes[1] and Neil J. Cronin[2]

1. Department for Physical Performance, Norwegian School of Sport Sciences, Norway
2. Neuromuscular Research Centre, Faculty of Sport and Health Sciences, University of Jyväskylä, Finland

**Abstract**
In vivo measurements of muscle architecture (i.e. the spatial arrangement of muscle fascicles) are routinely included in research and clinical settings to monitor muscle structure, function and plasticity. However, in most cases such measurements are performed manually, and more reliable and time-efficient automated methods are either lacking completely, or are inaccessible to those without expertise in image analysis.

In this work, we propose an ImageJ script to automate the entire analysis process of muscle architecture in ultrasound images: Simple Muscle Architecture Analysis (SMA). Images are filtered in the spatial and frequency domains with built-in commands and external plugins to highlight aponeuroses and fascicles. Fascicle dominant orientation is then computed in regions of interest using the *OrientationJ* plugin.

Bland-Altman plots of analyses performed manually or with SMA indicates that the automated analysis does not induce any systematic bias and that both methods agree equally through the range of measurements. Our test results illustrate the suitability of SMA to analyse images from superficial muscles acquired with a broad range of ultrasound settings.

**1. Introduction**
Musculoskeletal ultrasound imaging is used in a wide range of fields, including the study of muscle and tendon function [1,2], the effects of training on muscle architecture [3], and in the study of architectural parameters in different clinical populations [4]. The images generated by this method are complex, and require a great deal of time and effort from practitioners to interpret and extract meaning from them.

In recent years, efforts have been made to automate parts of this process [5–10]. However, these efforts have been fragmented, and suffer from a number of limitations: they often focus on analysing a single parameter of interest; most publications do not reveal specific details of how to implement the method; some methods rely on software that require expensive licence fees; the majority of methods are only semi-automated, requiring manual, subjective interpretation of at least some images; and often tracking methods involve complex mathematics and require computer programming experience. Collectively, these factors limit the widespread use of existing methods.

In this study we present Simple Muscle Architecture Analysis (SMA), a fully automated method of analysing muscle architectural parameters from individual images or collections of images. The method consists of a single macro in Fiji [11], which is a distribution of ImageJ [12,13], and is open source software that is commonly used to process ultrasound images. We provide full instructions for using the method, and no previous programming experience is required. In the following sections we present an overview of the method, as well as examples of analysis output and a range of test metrics.

**2. Materials and Methods**
*2.1. Overview*
In addition to the connective tissue surrounding bundles of muscle fibres (perimysium) and whole muscles (aponeuroses), ultrasound scans of muscle architecture acquired in brightness (B) mode show several types of echogenic tissues (e.g. fat, other types of connective tissue, blood vessels). A challenge associated with automatic segmentation of fascicles and aponeuroses is therefore to discriminate between signals from all echogenic regions (and artefactual echoes) and those from objects of interest. One of the originalities of the present approach is the implementation of filters in the frequency domain to isolate objects of interest (i.e. aponeuroses or fascicles) and improve their detection and registration.

The sample images used to test the present method were collected in previous projects, for which the subjects gave permission to use images by signing an informed consent and ethical approval was granted by the ethical committee of the Norwegian School of Sport Sciences.

The analysis method is summarised in Figure 1.

The analysis script is written in ImageJ 1.x macro language and runs in Fiji [11]. Although this scripting language is less powerful than others (e.g. Python), it is easy to learn and use, and was chosen with the goal of developing a tool customisable by most users. The script includes a header, a macro and auxiliary functions called within the macro. The header contains information about necessary plugins, the GNU General Public License, and the script parameters selectable by the user. The macro contains a main function *SingleImageAnalysis* and sub-scripts to display the results and run the macro on a single image or in batch mode (i.e. on series of images). The *SingleImageAnalysis* function is divided into four main sections, which perform the following operations: i) detect the ultrasound field of view (FoV), ii) segment and register superficial and deep aponeuroses, iii) measure dominant orientation of fascicles, and iv) calculate and compile variables of interest. The first three sections include a pre-processing step to isolate objects of interest and a function or a plugin (for fascicle orientation) to segment these objects. Several spatial filters are used in pre-processing. Although the purpose of their implementation is specified, their description would extend beyond the scope of this article. The reader is therefore referred to the relevant references for additional information.

*2.2. Detection of the ultrasound field of view*
Ultrasound images nearly always come with a frame containing various information about institution/patient name, scanning parameters and scaling. To proceed with the analysis of the ultrasound FoV, this section optionally (see the "Program description" section) identifies and crops the frame out. This step is achieved by filtering out most of the text from the frame by applying a convolution and a median filter, and binarizing the whole image with a median local threshold. The resulting segmentation enables the selection and deletion of the frame from a duplicate scan but does not alter the original image. The duplicate image of the FoV obtained at the end of this step is used in the three following sections.

*2.3. Detection of the aponeuroses*
The pre-processing of this section begins with a series of spatial filters using built-in commands or plugins (*Enhance Local Contrast, CLAHE* [14]; *Tubeness* [15]) in ImageJ/Fiji and a separately installed plugin (*Non-Local Means Denoising* [16,17]), to remove ultrasound noise and enhance muscle aponeuroses and fascicles. The *Tubeness* plugin is an implementation of multiscale vessel enhancement filtering first proposed by Frangi et al. [18]. The plugin scores each point on the basis of eigenvalues of the Hessian matrix, to determine an index of "tubeness". Although this approach was first developed to detect blood vessels, it can also effectively enhance muscle fascicles in ultrasound images [5]. A Fourier transform (FFT) is then applied to the filtered image. The resulting power spectrum is thresholded to retain the dominant signal from muscular aponeuroses and fascicles, and a mask is applied to the thresholded image to suppress features orientated in the expected direction of fascicles (see the "Program description" section). The inverse FFT of the filtered power spectrum is subsequently computed, and the edges of the aponeuroses are segmented (*Canny Edge Detector* plugin [19]) and registered using a custom-written function. This function uses a simple contrast threshold to search for parallel 'lines' in the horizontal direction, which typify the appearance of aponeuroses (see Figure 2B).

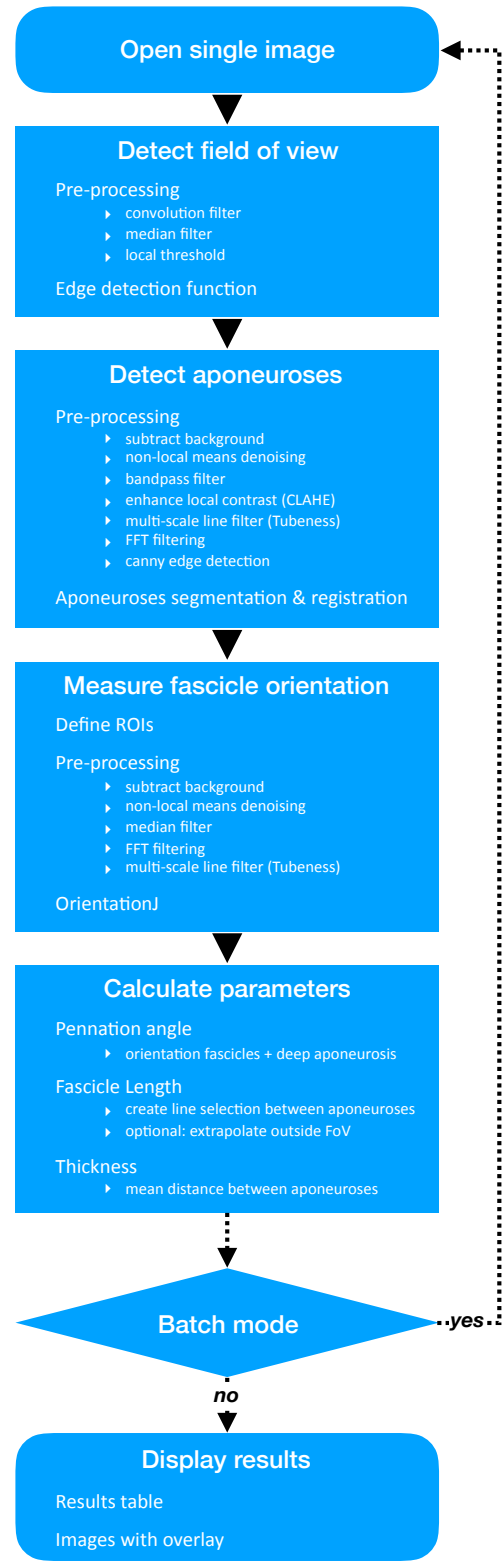

*Figure 1*: Workflow of the SMA (Simple Muscle Architecture Analysis) macro.

*2.4. Measurement of the dominant fascicle orientation*
At least two overlapping regions of interest (ROI) are defined along the width of non-filtered duplicates of the FoV, excluding aponeuroses (see the "Program description" section). Multiple overlapping ROIs were shown in pilot experiments to yield the dominant fascicle orientation more consistently than single or non-overlapping ROIs. Similarly to the aponeurosis detection, ROIs are first filtered spatially and in the frequency domain. However, to better preserve fascicular features, noise filtering is "lighter" than for aponeuroses (median filter and *Non-Local Means Denoising* plugin). Unless set manually (see the "Program description" section), a thresholding proportional to the brightness of the filtered image is applied to the power spectrum to mostly retain the frequencies corresponding to fascicles in the ROI. Following an inverse FFT, fascicles are enhanced with the *Tubeness* plugin. Their dominant orientation is measured within each ROI with the *OrientationJ* plugin [20,21]. *OrientationJ* is a plugin developed by Daniel Sage [20] to compute orientation and isotropy properties within a ROI, based on the evaluation of the gradient structure tensor in a local neighbourhood. The local window is characterized by a 2D Gaussian function of standard deviation σ, and here the gradient is computed with a (quasi-isotropic) cubic spline filter. The standard deviation of the Gaussian filter σ is defined by the user (see the "Program description" section), and should be proportional to the thickness of the fascicles (i.e. thicker fascicles are associated with a larger σ). The dominant angles from all ROIs are collected and either the greatest, the mean or the median angle - as selected by the user - is retained for each analysed image.

*2.5. Calculation of architecture parameters and display*
The angle of fascicle pennation is computed as the sum of the angles characterising the orientations of the deeper aponeurosis and fascicles, measured in previous steps. Fascicle length is computed as the length of a straight line running between aponeuroses, at an angle corresponding to the dominant fascicle orientation (either greatest, mean or median angle). Thus, the current version of the script estimates linear fascicle paths. A previous study comparing the length of gastrocnemius medialis fascicles when taking their curvature into account versus when assuming a straight path and parallel aponeuroses found a 6% difference during maximal contraction [22]. The difference seen in resting muscle was negligible. Because the current approach takes the orientation of aponeuroses into account, we expect an error negligible at rest and smaller than that

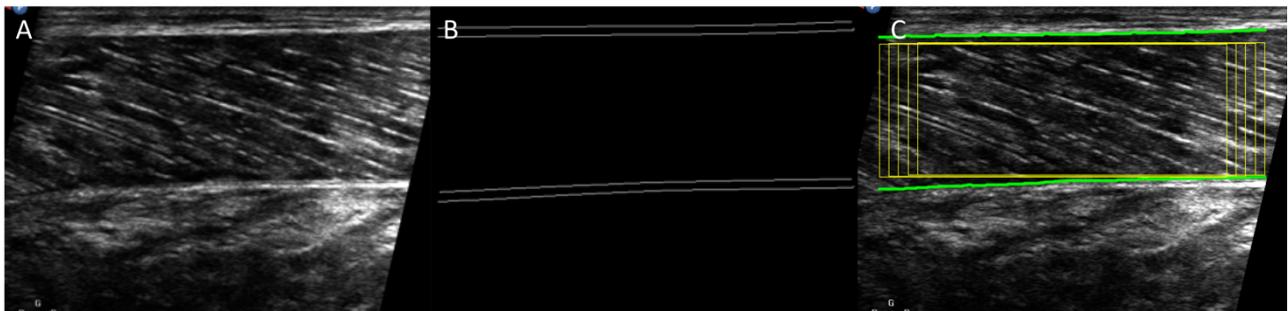

**Figure 2**: *Example of a raw image of the ultrasound field of view (A), aponeuroses edge detection (B) and ROIs (yellow) overlaid on the original image (C).*
*In C, the paths of the superficial and deep aponeuroses are also indicated as green lines.*

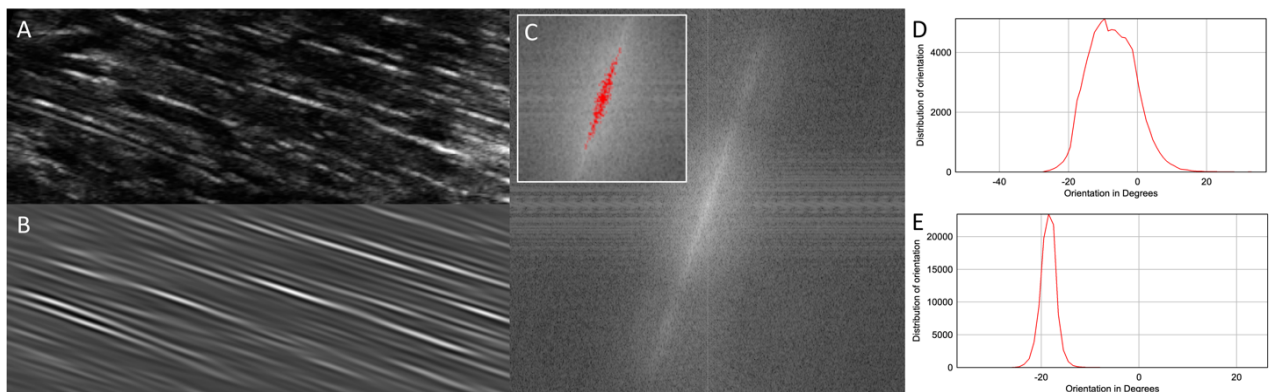

**Figure 3**: *Example of raw (A) and processed (B) images of fascicles.*

*Processing amplifies the signal of structures aligned with the direction of the fascicles by thresholding of the power spectrum (C). The resulting distribution of orientation computed with the OrientationJ plugin is narrower for the processed image (E) than for the original one (D).*

reported by Muramatsu and colleagues during contraction. Future implementations of the script may estimate curved paths. Although the analysis was developed for scans displaying entire fascicles, the script enables the extrapolation of aponeuroses if the composite fascicle runs outside the FoV. Muscle thickness is calculated as the mean distance between aponeuroses. The *SingleImageAnalysis* function ends after this step; if batch mode was selected the macro loops back to the beginning of the function and repeats the analysis for all other images. Numerical results are displayed in a table and a composite fascicular line is overlaid on corresponding original images, arranged in a stack.

*2.6. Program description*

The script can be opened and launched in Fiji without installation but end users are advised to install it in the Fiji.app/plugins folder. It can then be found in the Plugins menu of the Fiji software. When launched, the script opens a user interface divided into 5 sections. A brief description is displayed when the mouse pointer hovers over a parameter.

A message at the top reminds the user that the analysis script was designed for ultrasound scans displaying the proximal side to the left of the image, where fascicles would typically be angled upward to the left and the lower aponeurosis would be orientated horizontally or downwards to the left. The check box below this message allows the user to flip analysed images horizontally if required. In addition to the orientation of the image, we recommend following some basic scanning guidelines. The transducer should be used in the highest frequency range and the focal zone set to the deeper aponeurosis. When possible, the ultrasound beam can be steered towards the perpendicular to the direction of the fascicles, as long as aponeuroses still appear clearly. Using compound imaging is a preferred alternative to changing the beam orientation, when this feature is available. Any other feature (e.g. tissue harmonic imaging, speckle reduction) capable of reducing noise is also recommended. The dynamic range (or compression) should be moderate or low.

The following sections detail the choices offered in the user interface.

*2.6.1. Type of analysis*

The user chooses here to analyse single images, or, if multiple images are being analysed, selects target input and output folders. In the second case the extension corresponding to the file format must be specified. A check box enables the user to print all analysis parameters in the results table.

*2.6.2. Image cropping*

Images can be cropped manually. In this case the user is prompted to draw a selection around the field of view. In the case of folder analysis, the same dimensions of the cropping selection are used for all files, so it is advisable to only include images taken at the same scanning depth to avoid accidentally cropping the FoV. Manual cropping is recommended when aponeurosis detection fails, in particular when bright structures (e.g. other aponeuroses, bone edges) are visible under the muscle of interest.

*2.6.3. Aponeurosis detection*

The only parameter that can be adjusted at this step is the standard deviation - sigma - of the Gaussian used in the *Tubeness* plugin to convolve the image (see section 2.3. Detection of the aponeuroses). The default value is set to 10 on the basis of our sample images, but different spatial resolutions may require another value. A lower sigma (e.g. 8) is recommended when the aponeurosis edges are not detected accurately.

*2.6.4. Fascicle orientation*

In this section the user chooses the number and the size of the ROIs used to measure dominant fascicle orientation. These parameters define the degree of overlap and the depth of the ROIs. The choice of ROI height enables the computation of fascicle orientation within the whole image or in regions closer to the deeper aponeurosis. The latter is often favoured by researchers because of the better alignment between the deeper aponeurosis and the direction of muscle force, but this theoretical argument should be weighed against the smaller number and consistency of visible fascicles within smaller ROIs.

As explained in section *2.4. Measurement of the dominant fascicle orientation*, structures aligned in a different direction than the fascicles are filtered out by thresholding the power spectrum of ROIs. The threshold value is set automatically but the user can uncheck this option and set the threshold manually.

In the next step, the Laplacian of Gaussian filter value (referred to as σ in *2.4. Measurement of the dominant fascicle orientation*) can be set according to the spatial resolution of the image and expected fascicle thickness. The default value (4) was found to be suitable in most of our tests.

Finally, the user can select the method used to obtain the dominant fascicle orientation out of the values obtained in each ROI.

*2.6.5. Pixel scaling*

For single images or series of images acquired at the same depth, measurements can be scaled if this option is checked. The user should then select the scanning depth, and at the onset of the analysis when prompted, draw a straight line over the scale bar usually included in the frame of ultrasound scans (as is normally done when scaling images in ImageJ/Fiji).

*2.6.6. System requirements*

Hardware/software specifications are based on Fiji/ImageJ requirements. Currently, the following systems are supported:
- Windows XP, Vista, 7, 8 and 10
- Mac OS X 10.8 "Mountain Lion" or later
- Linux on amd64 and x86 architectures

However, the capability to use Java 8 runtime is the only real requirement.

### 2.6.7. Mode of availability

The SMA script and the necessary (*Non Local Means Denoise*, *Canny Edge Detector* and *OrientationJ*) or optional (*FFMPEG*) plugins are installed via the ImageJ Updater interface, by adding relevant update sites. This relatively simple procedure has the advantage of automatically updating all of the required components and libraries. Generic steps for adding an update site are explained on the ImageJ website[1], but detailed instructions for installing SMA are provided in Appendix A of the supplementary material.

## 3. Results

The analysis is currently optimised for superficial muscle architecture images taken at rest, with entire fascicles visible within the FoV. However, aponeuroses can be extrapolated linearly and the length of elongated fascicles (e.g. when the muscle is being stretched) can be estimated with satisfactory accuracy. The primary purpose of the macro is therefore to obtain reliable estimates of muscle architecture in repeated measurements from the same individuals.

In the first example below, we compared the reliability and validity of this type of analysis against manual analysis (Figures 4 & 5). To illustrate the effect of variations in image parameters due to the use of different equipment and ultrasound settings, sample scans of gastrocnemius medialis muscle were acquired from 10 individuals (Sample A), at 9 MHz, with a 96-element transducer (60mm, LV7.5/60/96Z, LogicScan 128 EXT-1Z, Telemed, Lithuania), and 10 different sample scans (Sample B) were acquired at 12 MHz, with a 128-element transducer (50 mm, 5–12 MHz HD11XE, Phillips, Bothell, Washington, USA). Ultrasound parameters were adjusted to visualise fascicles and aponeuroses appropriately but differently for each sample (e.g. contrast, smoothing), to reflect a broader range of image quality than obtained with hardware alone. Therefore, the results of the analysis of these samples do not reflect brand-specific capabilities of the ultrasound systems used here. Both samples were analysed with the SMA script, and manually (and independently) by the two authors. Manual analysis consisted of digitising 3 fascicles per image and the distance between superficial and deep aponeuroses at 3 different sites (approximately 25, 50 and 75% of image width). An average of these measurements was used to represent pennation angle, fascicle length and muscle thickness.

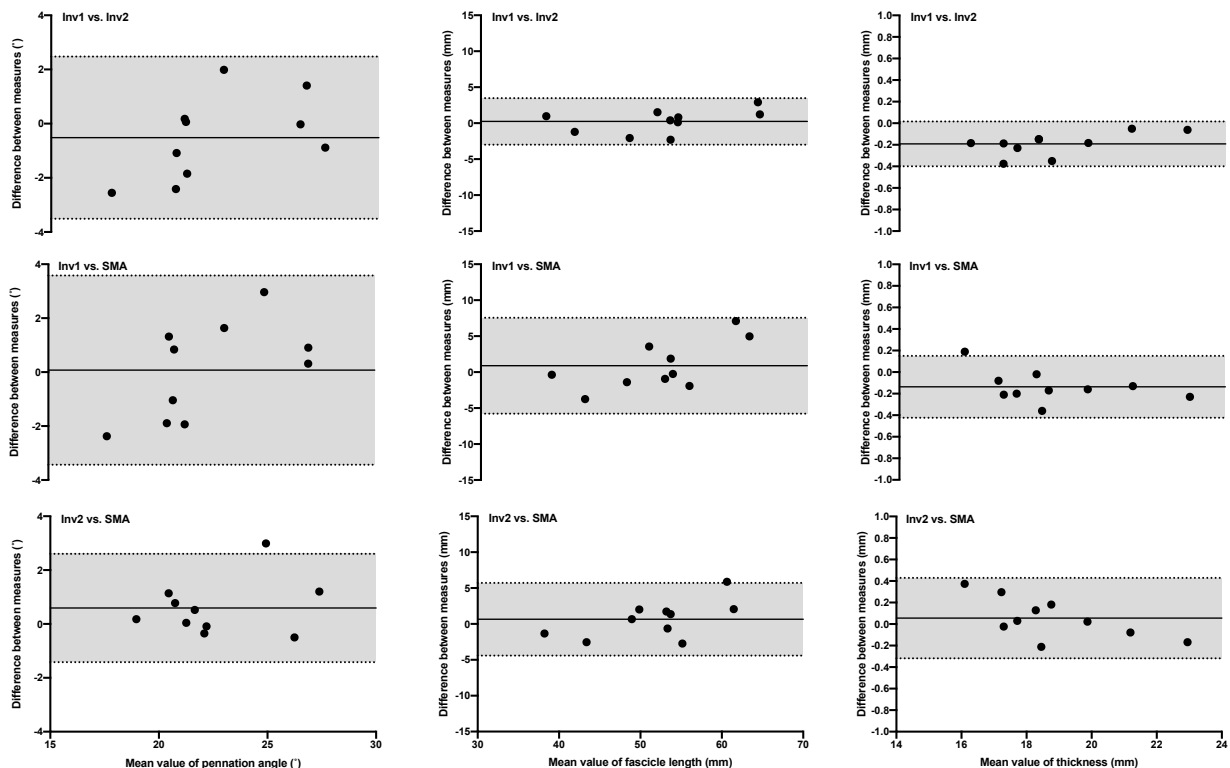

*Figure 4*: Bland-Altman plots of pennation angle, fascicle length and muscle thickness obtained from scans of the gastrocnemius medialis in Sample A.
*Comparisons between automated analysis (SMA), manual analyses from investigator 1 (Inv1) and investigator 2 (Inv2) are illustrated by differences between pairs of measurements as a function of the mean measurements. Solid and dotted lines depict bias and 95% limits of agreement, respectively.*

---

[1] https://imagej.net/Following_an_update_site

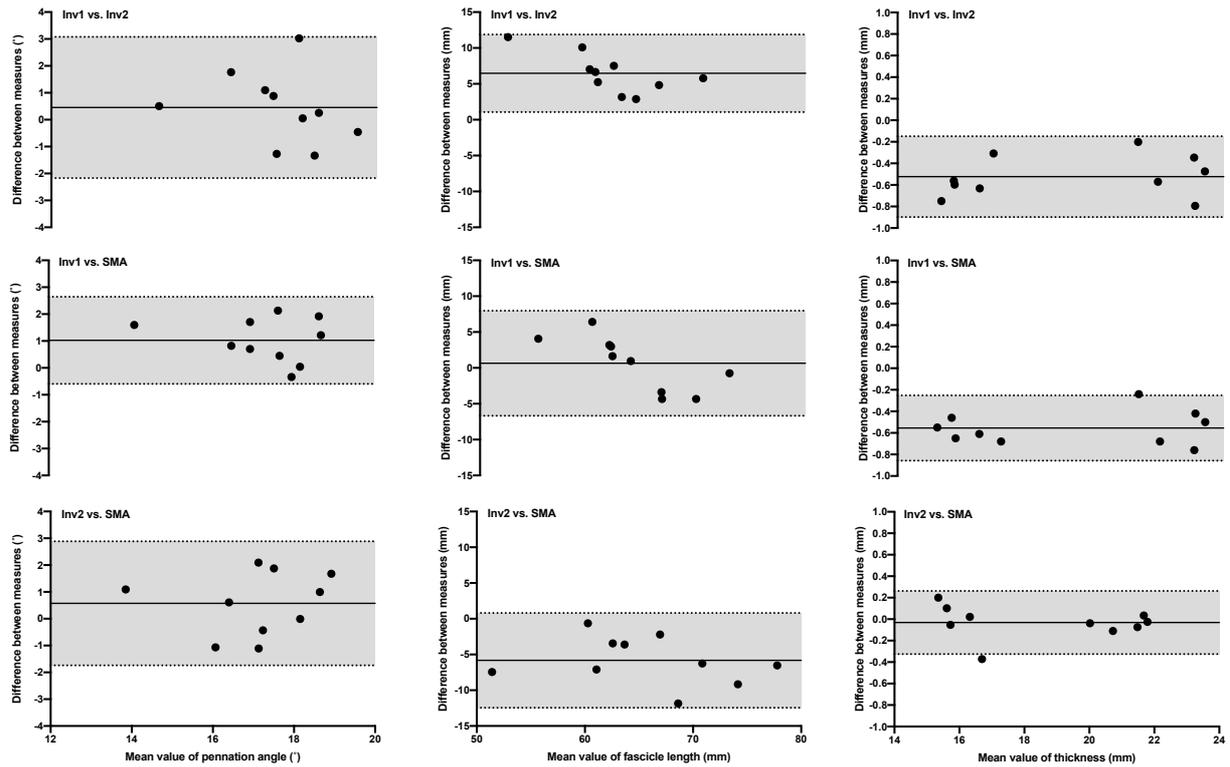

*Figure 5*: Bland-Altman plots of pennation angle, fascicle length and muscle thickness obtained from scans of the gastrocnemius medialis in Sample B.
*Comparisons between automated analysis (SMA), manual analyses from investigator 1 (Inv1) and investigator 2 (Inv2) are illustrated by differences between pairs of measurements as a function of the mean measurements. Solid and dotted lines depict bias and 95% limits of agreement, respectively.*

Tracking of architecture parameters in image series is also possible with the SMA script, as long as fascicles are imaged consistently, as in during slow contractions or passive stretching. If scans were acquired as movies, the user should first install FFMPEG plugins [23] to be able to import them into Fiji, before saving them as image sequences. An example of analysis of images (Sample C) acquired at 12 MHz, with a 128-element transducer (60mm, HL9.0/60/128Z-2, LogicScan 128 EXT-1Z, Telemed, Lithuania) on the tibialis anterior muscle of an individual is illustrated in Figure 6. Scans were acquired at 15 frames per second, during an unconstrained dorsiflexion-plantarflexion movement. The analysis was performed with the following parameters: automatic cropping, Tubeness sigma (aponeurosis detection) = 7, 3 ROIs, ROI width = 60, ROI height = 90, automatic thresholding of the ROIs power spectrum, OrientationJ σ = 4 and maximal value of all dominant orientations detected in ROIs. The mean analysis time per frame was 8.4s on the computer used for this example.

The three datasets analysed during the current study are available as separate compressed files at the following address:
https://data.mendeley.com/datasets/dpmf9bz8pt/1.

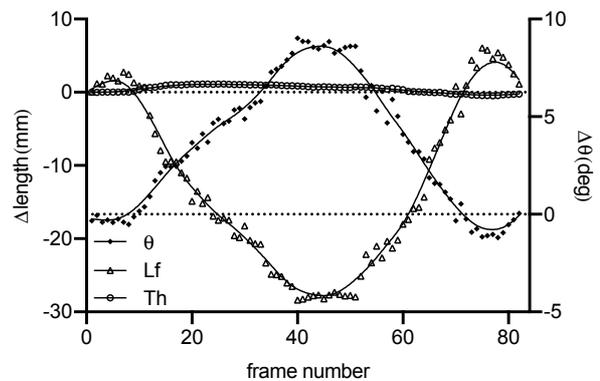

*Figure 6*: Changes in thickness (Th), fascicle length (Lf) and pennation angle (Θ) in the tibialis anterior of one individual during unrestrained dorsiflexion and plantarflexion (Sample C). Raw data were fitted with a lowess curve to improve visualisation of change patterns.

## 4. Discussion

The measurements of muscle architecture parameters obtained with SMA are within the range of expected values obtained manually. Bland-Altman plots show that the automated analysis does not induce any substantial bias, in particular when taking the inter-rater comparison into account. The bias in pennation angle is in all cases less than or equal to 1°. Image quality (i.e. spatial resolution and contrast) probably influenced the estimation of fascicle length, with a bias of up to 6.5 mm in Sample A and less than 1 mm in Sample B. Importantly, the bias magnitude was not specific to manual vs. automated analysis. The bias in muscle thickness was negligible in all comparisons (<1 mm). None of the measured biases were found to be proportional to averaged values, indicating that manual and automatic analyses agree equally through the range of measurements. The validity of muscle architecture measurements performed manually has previously been shown [24]. The present comparisons demonstrate that the results obtained with SMA and manual measurements are comparable and, indirectly, equally valid. In addition, automated measurements improve the reliability of muscle architecture measurements by removing the variability induced by manual segmentation, particularly when multiple raters are involved [25].

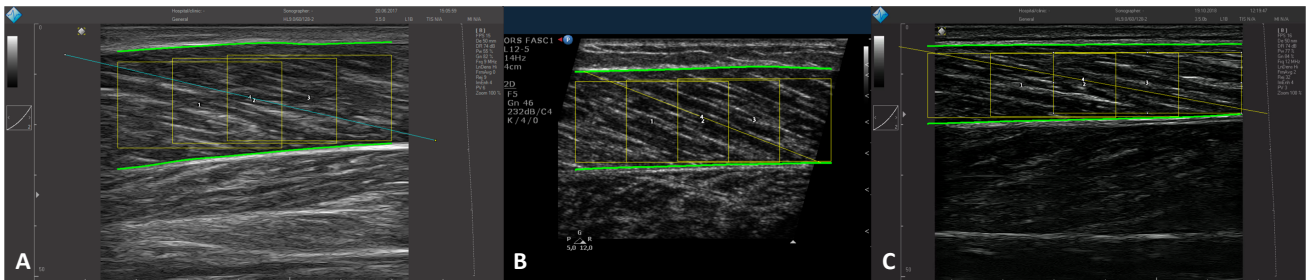

*Figure 7*: Examples of processed scans from Samples A, B and C, with overlay of composite fascicles and regions of interest used in the analysis.

As shown in examples of scans processed with the SMA script (Figure 7), the implemented filtering and user adjustments allow the processing of scans acquired with a broad range of ultrasound settings.

However, the analysis relies heavily on sufficient contrast and homogeneity of grey values when imaging fascicles and aponeuroses. For instance, large differences between grey values of the lower and upper aponeuroses may cause the detection to fail. Likewise, inconsistent grey values along the length of an aponeurosis may affect its segmentation. Users are advised to test the suitability of new image settings prior to data collection. This step is important since different settings will influence the results of any analysis. For instance, different *sigma* values chosen to detect the aponeurosis (see sections 2.3 and 2.6.3) will yield slight differences in the computed orientation of these structures. Similarly, the parameters used to compute the dominant fascicle orientation (e.g. ROI size and number, σ, see section 2.6.4) will influence the results. In addition to testing these parameters on sample data, the user is advised to keep the same settings when analysing subsequently collected data. Except in cases where parts of the image need to be excluded (e.g. due to excessive connective tissue) and cropping must be done manually, the above precautions will generally enable the analysis process to be fully automated if desired. Finally, the script was designed for superficial muscles. Deeper muscles may be targeted by manually cropping the FoV but the above requirements may limit their analysis.

5. Conclusion

The automated analysis proposed here yields results comparable to those obtained manually. The SMA script, the first implementation of such an analysis to be proposed as a free and open source software, is therefore a suitable tool to reduce some of the variability introduced by manual analyses, and to improve the overall quality of muscle architecture measurements. Since it relies on sufficient contrast and homogeneity of grey values, it is mainly intended for use with images from resting muscles, but our test results also show that scans acquired during motion can be analysed, as long as qualitative image criteria are met.

References
1. Cronin, N.J.; Lichtwark, G. a The use of ultrasound to study muscle-tendon function in human posture and locomotion. *Gait & posture* **2012**.
2. Seynnes, O.R.; Bojsen-Møller, J.; Albracht, K.; Arndt, A.; Cronin, N.J.; Finni, T.; Magnusson, S.P. Ultrasound-based testing of tendon mechanical properties: a critical evaluation. *Journal of Applied Physiology* **2015**, *118*, 133–141.
3. Timmins, R.G.; Shield, A.J.; Williams, M.D.; Lorenzen, C.; Opar, D.A. Architectural adaptations of muscle to training and injury: a narrative review outlining the contributions by fascicle length, pennation angle and muscle thickness. *Br J Sports Med* **2016**.
4. Connolly, B.; MacBean, V.; Crowley, C.; Lunt, A.; Moxham, J.; Rafferty, G.F.; Hart, N. Ultrasound for the assessment of peripheral skeletal muscle architecture in critical illness: a systematic review. *Crit. Care Med.* **2015**, *43*, 897–905.
5. Rana, M.; Hamarneh, G.; Wakeling, J.M. Automated tracking of muscle fascicle orientation in B-mode ultrasound images. *Journal of biomechanics* **2009**, *42*, 2068–73.
6. Cronin, N.J.; Carty, C.P.; Barrett, R.S.; Lichtwark, G. Automatic tracking of medial gastrocnemius fascicle length during human locomotion. *J. Appl. Physiol.* **2011**, *111*, 1491–1496.
7. Zhou, G.-Q.; Chan, P.; Zheng, Y.-P. Automatic Measurement of Pennation Angle and Fascicle Length of Gastrocnemius Muscles Using Real-time Ultrasound Imaging. *Ultrasonics* **2014**.
8. Farris, D.J.; Lichtwark, G.A. UltraTrack: Software for semi-automated tracking of muscle fascicles in

sequences of B-mode ultrasound images. *Comput Methods Programs Biomed* **2016**, *128*, 111–118.
9. Caresio, C.; Salvi, M.; Molinari, F.; Meiburger, K.M.; Minetto, M.A. Fully Automated Muscle Ultrasound Analysis (MUSA): Robust and Accurate Muscle Thickness Measurement. *Ultrasound in Medicine & Biology* **2017**, *43*, 195–205.
10. Nikolaidou, M.E.; Marzilger, R.; Bohm, S.; Mersmann, F.; Arampatzis, A. Operating length and velocity of human M. vastus lateralis fascicles during vertical jumping. *Royal Society Open Science* **2017**, *4*, 170185.
11. Schindelin, J.; Arganda-Carreras, I.; Frise, E.; Kaynig, V.; Longair, M.; Pietzsch, T.; Preibisch, S.; Rueden, C.; Saalfeld, S.; Schmid, B.; et al. Fiji: an open-source platform for biological-image analysis. *Nat. Methods* **2012**, *9*, 676–682.
12. Schneider, C.A.; Rasband, W.S.; Eliceiri, K.W. NIH Image to ImageJ: 25 years of image analysis. *Nat. Methods* **2012**, *9*, 671–675.
13. Rueden, C.T.; Schindelin, J.; Hiner, M.C.; DeZonia, B.E.; Walter, A.E.; Arena, E.T.; Eliceiri, K.W. ImageJ2: ImageJ for the next generation of scientific image data. *BMC Bioinformatics* **2017**, *18*, 529.
14. Saalfeld, S. Enhance Local Contrast (CLAHE) - ImageJ Available online: https://imagej.net/Enhance_Local_Contrast_(CLAHE) (accessed on Oct 17, 2018).
15. Longair, M.; Preibisch, S. Tubeness - ImageJ Available online: https://imagej.net/Tubeness (accessed on Oct 17, 2018).
16. Buades, A.; Coll, B.; Morel, J.-M. Non-Local Means Denoising. *Image Processing On Line* **2011**, *1*, 208–212.
17. Darbon, J.; Cunha, A.; Chan, T.F.; Osher, S.; Jensen, G.J. Fast nonlocal filtering applied to electron cryomicroscopy. In Proceedings of the 2008 5th IEEE International Symposium on Biomedical Imaging: From Nano to Macro; 2008; pp. 1331–1334.
18. Frangi, A.F.; Niessen, W.J.; Vincken, K.L.; Viergever, M.A. Multiscale vessel enhancement filtering. In Proceedings of the Medical Image Computing and Computer-Assisted Intervention — MICCAI'98; Wells, W.M., Colchester, A., Delp, S., Eds.; Springer Berlin Heidelberg, 1998; pp. 130–137.
19. Gibara, T. Canny Edge Detector Available online: https://imagej.nih.gov/ij/plugins/canny/index.html (accessed on Oct 17, 2018).
20. Püspöki, Z.; Storath, M.; Sage, D.; Unser, M. Transforms and Operators for Directional Bioimage Analysis: A Survey. In *Focus on Bio-Image Informatics*; De Vos, W.H., Munck, S., Timmermans, J.-P., Eds.; Advances in Anatomy, Embryology and Cell Biology; Springer International Publishing: Cham, 2016; pp. 69–93 ISBN 978-3-319-28549-8.
21. Fonck, E.; Feigl, G.G.; Fasel, J.; Sage, D.; Unser, M.; Rüfenacht, D. a; Stergiopulos, N. Effect of aging on elastin functionality in human cerebral arteries. *Stroke; a journal of cerebral circulation* **2009**, *40*, 2552–6.
22. Muramatsu, T.; Muraoka, T.; Kawakami, Y.; Shibayama, A.; Fukunaga, T. In vivo determination of fascicle curvature in contracting human skeletal muscles. *Journal of Applied Physiology* **2002**, *92*, 129–134.
23. Schindelin, J. FFMPEG plugins Available online: http://fiji.sc/~schindelin/ffmpeg-plugins/ (accessed on Nov 1, 2018).
24. Kwah, L.K.; Pinto, R.Z.; Diong, J.; Herbert, R.D. Reliability and validity of ultrasound measurements of muscle fascicle length and pennation in humans: a systematic review. *J. Appl. Physiol.* **2013**, *114*, 761–769.
25. König, N.; Cassel, M.; Intziegianni, K.; Mayer, F. Inter-rater reliability and measurement error of sonographic muscle architecture assessments. *J Ultrasound Med* **2014**, *33*, 769–777.

# Appendix A

**Installation instructions for SMA (Simple Muscle Architecture Analysis): An ImageJ/Fiji based macro for automated analysis of B-mode ultrasound images.**

Step 1: Download the Fiji software using the following link. **Please read the warning on this page about where to install Fiji, as this can affect your ability to get updates**:
https://imagej.net/Fiji/Downloads

Step 2: After downloading and running Fiji, you need to gather a few dependencies. First, click *Help* from the dropdown menu in Fiji, then choose *Update*. In the Updater window, select *Manage update sites*. In the window that appears, you will notice that some boxes are already ticked. This means that if/when an update is released for the corresponding programme, your machine will automatically install it at the first opportunity.

Scroll down the list and tick the *BIG-EPFL* and *Biomedgroup* update sites. Then, click the *Add update site* button and scroll to the bottom of the list, where a new update site has been added. Modify the details of the new site as follows (to modify a section, double-click it):
Name: SMA
URL: http://sites.imagej.net/SMA
Host: webdav:SMA

After entering the details, make sure that the box to the left of the SMA text is ticked. Click *Close* once you have finished editing. Click *Apply changes* to confirm the installation. Close Fiji and re-open it. You are now ready to use SMA.

Optional step 3 (recommended): It is convenient to add the SMA macro to the list of available plugins so that you can run SMA from the dropdown menu. Otherwise you'll need to open the SMA_1_5X.ijm file from the *Fiji macros* folder, and run it manually.

On Mac, SMA would automatically be added to the *Plugins* dropdown list. To add SMA to the *Plugins* dropdown list on Windows, select *Plugins* from the dropdown menu, then *Install*, and choose the SMA_1_5X.ijm file (the X will depend on what version you have), which you can find by double-clicking your main *Fiji* folder, then double-clicking the *macros* folder, which should contain the target file. Select it and choose *open*. Finally, you will see a different window asking where you would like to save this file. Choose the Fiji plugins folder (should be the default) and click *save*. The SMA macro should now appear at the bottom of the dropdown *Plugins* list.

Optional step 4: To analyse movie frames (see limitations in the companion article) with SMA, movies must first be imported into Fiji and converted into image sequences (NB: movies can also be converted in advance using some other software). This is possible if the update site for the *FFMEG* plugins is added (follow the procedure described above).

A movie file can then be imported by clicking *File > Import > Movie (FFMPEG)*, selecting the movie and accepting the default *Import options*. Once imported, the movie can be converted into an image sequence in several ways.
One way is to click *Image > Stacks > Tools > Make Substack…* and then enter the range of frames to be exported. The obtained image stack can then be archived by clicking *File > Save as >Image Sequence*, and selecting the appropriate file format and saving location.